\documentclass[prc,twocolumn,showpacs,amsmath,amssymb,floatfix]{revtex4-1}
\usepackage{graphicx}
\usepackage{supertabular}
\usepackage{epsfig}
\usepackage{dcolumn}
\usepackage{bm}
\usepackage{wrapfig} 
\usepackage{amsmath}
\usepackage{amsfonts}
\usepackage{mathrsfs}
\begin{document} 

\title{Microscopic description of the ground state properties of  
recently reported new isotopes}
\author{H. M. Devaraja$^{1}$, Y. K. Gambhir$^{1,2}$, A. Bhagwat$^{3}$, 
M. Gupta$^{1}$, S. Heinz$^{4,5}$, G. M\"unzenberg$^{1,4}$}
\affiliation{
$^{1}$ Manipal Centre for Natural Sciences, Manipal University, 
Manipal 576014, Karnataka, India \\
$^{2}$ Department of Physics, IIT Bombay, Powai, Mumbai 400076, India \\
$^{3}$ UM-DAE Centre for Excellence in Basic Sciences, Mumbai 400 098, India \\
$^{4}$ GSI Helmholtzzentrum fur Schwerionenforschung GmbH, 64291 Darmstadt, 
Germany \\
$^{5}$ Justus-Liebig-Universitat Giessen, II, Physikalisches Institut, 
35392 Geissen, Germany}
\date{\today}

\begin{abstract}
Microscopic investigations for the observed properties of the recently reported 
five unstable new isotopes are carried out. The ground state properties are 
calculated in the relativistic mean field (RMF) framework and the results  
reproduce the experiment well as expected. The  $\alpha$ - decay lifetimes 
are calculated in the double folding model using WKB approximation which 
requires the relevant $Q$ – values of $\alpha$ - decay and the $\alpha$ - daughter
  ($V_{\alpha  D}$)  potential. 
The latter is obtained by folding the effective M3Y nucleon – nucleon 
potential with the RMF nucleon density distributions for the daughter nucleus 
and that of the $\alpha$ particle which is assumed to be of Gaussian shape. the corresponding 
decay half - lives obtained by using available phenomenological expression are also presented, 
discussed and compared.
 It is  observed that  the  use of accurate  $Q$ – values ( very close to the  experimental values) 
 is crucial for the  reliable description of the experimental   $\alpha$ - decay half-lives in the WKB framework. 

\end{abstract}
\maketitle

\section{Introduction}
Recently, five new neutron deficient isotopes with  $Z\ge92$ have been 
 observed in multi-nucleon transfer reactions \cite{DEV.15}. 
The  decay energies (E$_\alpha$) and half-lives of the respective 
$\alpha$ - decay chains of these new isotopes, $^{216}_{\,\,92}$U, $^{223}_{\,\,95}$Am,
$^{219}_{\,\,93}$Np, $^{229}_{\,\,95}$Am and $^{233}_{\,\,97}$Bk have been measured.  
The corresponding Q (also denoted by Q$_\alpha$) values are trivially extracted using E$_\alpha$.
The nuclei appearing in these decay chains have odd nuclear mass number A, 
except for the nuclei appearing in the $\alpha$ – decay chain of  $^{216}_{\,\,92}$U. 
The new isotopes were the products of multi-nucleon transfer reactions in collisions of
 $^{48}$Ca + $^{248}$Cm carried out at the velocity filter SHIP \cite{MUN.79} at GSI. These 
decay chains are shown in Fig. 1. 

\begin{figure*}
\includegraphics[width=16cm,height=8cm]{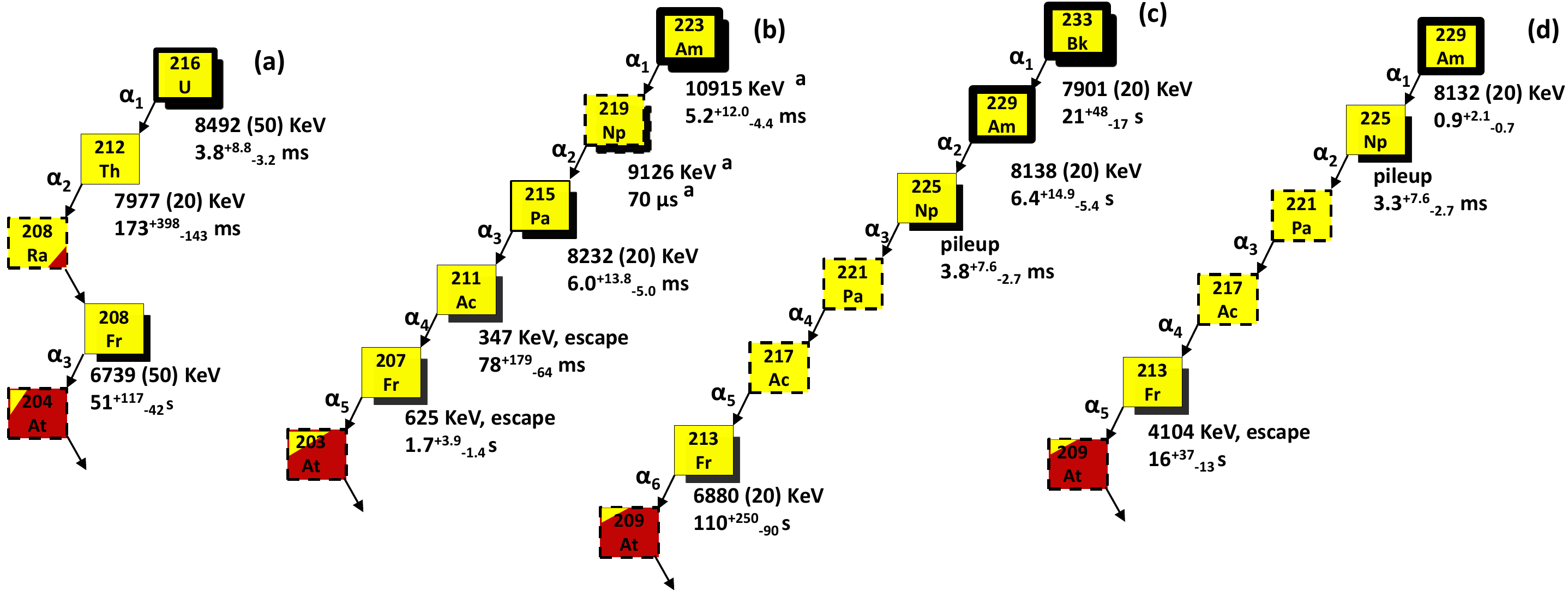}
\caption{Decay chains of the 
newly reported isotopes \cite{DEV.15}, observed in multi-nucleon transfer reactions of 
$^{48}$Ca + $^{248}$Cm. The squares framed with full lines mark isotopes which 
were observed in the experiment while dashed frames indicate isotopes which 
are expected members of the respective decay chain but were not observed. 
The new isotopes are marked by squares with bold frames. Events marked with a shadow were
 registered during the beam-off periods. The extracted Q$_\alpha$ values and half-lives ($T_{1/2}$) 
are given for each observed 
nucleus.}
\label{fig1}
\end{figure*}

In this brief communication we are mainly concerned with the calculation of $\alpha$ - decay half-lives for the
nuclei appearing in $\alpha$ - decay chains of these new isotopes.
The calculation requires the relevant $Q$ – values of $\alpha$ - decay and   in addition the $\alpha$ - daughter 
potential  for  the  WKB calculations. For completeness,  we first calculate
 the ground state properties of 
these new isotopes along with the nuclei appearing in their  $\alpha$-decay 
chains  using  the relativistic mean field (RMF) framework 
\cite{PGR.89,YKG.90,PRI.96,VRE.05,BHA.11} which is akin to the energy density 
functional formalism. The calculated total binding energies, radii and sizes, the deformation parameter
 and $Q$ – values of the $\alpha$ - decay chains  are briefly discussed.  The $\alpha$ - decay half-lives are then calculated 
in the WKB framework employing  $V_{A-D}$  potential obtained using double folding model ($t {\rho \rho }$ 
approximation). 
The results are discussed in detail and are compared with the corresponding experimental values and those 
obtained by using the  recent phenomenological expression.

The details of the calculation are briefly described  in Section 2. The results are discussed in Section 3, while the 
summary and conclusion are reserved for Section 4.

\section{Details of the calculation}
The RMF \cite{PGR.89,YKG.90,PRI.96,VRE.05,BHA.11} starts with a Lagrangian 
describing Dirac spinor nucleons interacting via the meson and the 
electromagnetic photon fields. In the mean field approximation, along  with 
time reversal invariance, for the static case one ends up with a  nonlinear 
set  of coupled equations - the Dirac type equations with potential terms 
involving the meson and photon fields describing the  nucleon dynamics and 
Klein - Gordon type equations for mesons with sources involving the nucleon 
current and densities. This set of equations is to be solved self-consistently. 
The pairing correlations are incorporated either through the 
conventional Bardeen-Cooper-Schrieffer (BCS) type procedure (constant gap 
approximation) for the 
calculation of the occupation probabilities, which gives rise to a set of 
equations  known as the relativistic mean field (RMF) equations or self 
consistently resulting into a set of well known relativistic Hartree-Bogoliubov 
(RHB) equations.  The pairing field $ \hat{\Delta}$  appearing in the RHB equations becomes
diagonal in the constant gap approximation and so decouples into a set of diagonal
 matrices resulting in the BCS type expressions for the occupation probabilities. 
As a result the  RHB equations reduce to the corresponding RMF equations for the occupation 
probabilities with constant pairing gap ${\Delta}$ .

\begin{table*}[htb]
\begin{center}
\caption{Calculated ground state properties of the nuclei relevant to this work.}
\addtolength{\tabcolsep}{3pt}
\begin{tabular}{ccccccccccc} \hline
            & \multicolumn{4}{c}{BE (MeV)} & \multicolumn{2}{c}{$r_p$ (fm)} & \multicolumn{2}{c}{$r_n-r_p$ (fm)} &  $\beta$ & Tagged State \\
\cline{2-5}\cline{6-7}\cline{8-9}
Nucleus                      &  NL3*  & PC-PK1 &  AB   &  AU   &  NL3* & PC-PK1 &  NL3* & PC-PK1 & (NL3*) & (Proton)      \\\hline \vspace{2pt}
$^{233}_{\,\,97}$Bk$_{136}$  &1757.75 &1744.13 &1751.93&       & 5.805 &  5.795 & 0.163 & 0.132  & 0.218 &   \\\vspace{3pt}
$^{229}_{\,\,95}$Am$_{134}$  &1737.42 &1726.90 &1731.93&1731.86$^{+}$ & 5.741 &  5.749 & 0.170 & 0.145  & 0.059 &   \\\vspace{3pt}
$^{225}_{\,\,93}$Np$_{132}$  &1718.32 &1709.89 &1711.16&1711.70& 5.691 &  5.702 & 0.185 & 0.158  & 0.043 &   \\\vspace{3pt}
$^{221}_{\,\,91}$Pa$_{130}$  &1700.15 &1692.80 &1692.64&1692.20& 5.640 &  5.651 & 0.198 & 0.172  & 0.028 &   \\\vspace{3pt}
$^{217}_{\,\,89}$Ac$_{128}$  &1680.80 &1674.84 &1673.50&1673.14& 5.589 &  5.598 & 0.209 & 0.183  &-0.008 &   \\\vspace{3pt}
$^{213}_{\,\,87}$Fr$_{126}$  &1662.68 &1657.61 &1654.80&1654.68& 5.537 &  5.544 & 0.218 & 0.195  & 0.000 &   \\\vspace{3pt}
$^{209}_{\,\,85}$At$_{124}$  &1638.35 &1633.88 &1633.13&1633.24& 5.496 &  5.502 & 0.226 & 0.202  &-0.008 &   \\\vspace{3pt}
                             &        &        &       &       &       &        &       &        &       &   \\ \vspace{3pt}
$^{223}_{\,\,95}$Am$_{128}$  &1693.68 &1683.99 &1682.76&       & 5.679 &  5.682 & 0.138 & 0.121  & 0.004 &   \\\vspace{3pt}
$^{219}_{\,\,93}$Np$_{126}$  &1677.10 &1669.32 &1665.28&1665.50$^{+}$ & 5.628 &  5.634 & 0.150 & 0.133  & 0.002 &   \\\vspace{3pt}
$^{215}_{\,\,91}$Pa$_{124}$  &1656.79 &1649.14 &1646.04&1646.27& 5.591 &  5.598 & 0.158 & 0.137  & 0.021 &   \\\vspace{3pt}
$^{211}_{\,\,89}$Ac$_{122}$  &1634.62 &1626.52 &1626.18&1626.21& 5.554 &  5.559 & 0.164 & 0.143  &-0.039 &   \\\vspace{3pt}
$^{207}_{\,\,87}$Fr$_{120}$  &1616.67 &1603.36 &1605.49&1605.54& 5.515 &  5.518 & 0.171 & 0.150  &-0.055 &   \\\vspace{3pt}
$^{203}_{\,\,85}$At$_{118}$  &1588.48 &1580.23 &1584.01&1584.14& 5.475 &  5.474 & 0.182 & 0.159  &-0.073 &   \\\vspace{3pt}
                             &        &        &       &       &       &        &       &        &       &   \\ \vspace{3pt}
$^{216}_{\,\,92}$U$_{124}$   &1658.88 &1651.32 &1647.86&       & 5.604 &  5.613 & 0.149 & 0.127  & 0.000 &   \\\vspace{3pt}
$^{212}_{\,\,90}$Th$_{122}$  &1636.64 &1629.12 &1628.37&1628.61& 5.569 &  5.575 & 0.153 & 0.132  &-0.017 &   \\\vspace{3pt}
$^{208}_{\,\,88}$Ra$_{120}$  &1614.50 &1606.22 &1608.05&1608.27& 5.532 &  5.534 & 0.160 & 0.140  & 0.061 &   \\\vspace{3pt}
                             &        &        &       &       &       &        &       &        &       &   \\ \vspace{3pt}
$^{208}_{\,\,87}$Fr$_{121}$  &1620.74 &1612.13 &1613.25&1613.44& 5.519 &  5.523 & 0.179 & 0.157  &-0.050 &   \\\vspace{3pt}
$^{204}_{\,\,85}$At$_{119}$  &1597.12 &1588.53 &1591.81&1591.93& 5.478 &  5.479 & 0.189 & 0.166  &-0.060 &   \\ \hline 
\end{tabular}
\end{center}
\end{table*}

Here, we employ the well established basis expansion method for the solution of 
the RMF/RHB equations.  We choose spherical (axially deformed) Harmonic 
Oscillator (HO) basis for the spherical (deformed) nuclei.  Accordingly, 
we expand the nucleon spinors and separately also the meson fields in this 
basis. The e.m. photon field is treated in the conventional manner. The 
details are given in Ref. \cite{YKG.90}. Explicit calculations require the 
parameters appearing in the Lagrangian, namely  the nucleon and meson masses 
and their coupling constants together with the additional parameters like the 
non-linear terms in which the isoscalar - scalar sigma meson is assumed to 
move. These parameters are determined through a $\chi^2$ fit to the observed 
ground state properties of a few selected spherical nuclei. This parameter 
set is then frozen and is used for all the nuclei spread over the entire 
nuclear chart including super heavy elements (SHE). The parameter set so 
obtained is not unique and depends upon the detailed ground state properties 
included in the fit. Several such parameter sets are available in the 
literature (see, for example, Ref. \cite{AGB.14}). Most of the recently reported
 parameter
sets are mainly used for the calculation of the even - even (e - e) nuclei.
The calculated results 
for the nuclear ground state properties obtained with these different 
parameter sets  are qualitatively  similar (e.g. see \cite{AGB.14,ZHA.05}),   
some differences do appear at a finer level.  Specifically, the use of recently reported
 parameter sets does improve the calculated binding energies and the Q$_\alpha$ values significantly. 
However, the resulting nuclear 
radii (sizes) and the density distributions remain almost identical.  

We employ here the  Lagrangian parameter set  NL3*  \cite{LAL.09} 
(the improved version 
of the most widely used in the past,  the  set NL3 \cite{LAL.97}) in our illustrative 
calculations.
For the calculation of partial occupation probabilities arising due to 
important pairing correlations, we use here the finite range Gogny-D1S 
interaction \cite{BER.84,GON.96} which is known to have the right content 
of pairing, while solving the RHB equations in the spherical H.O. basis. To 
determine the proton and neutron pairing gaps, we calculate the respective 
pairing energies in the spherical RMF  such that  the corresponding pairing 
energies obtained in the RHB with Gogny-D1S pairing interaction are reproduced. 
These gaps are then also used  in the  RMF code for deformed nuclei.  

The nuclei under consideration in the present work, have odd mass number 
A except for the nuclei appearing in the alpha decay chain of $^{216}_{\,\,92}$U.
 It is to be mentioned  that for odd/odd-odd deformed nuclei the time reversal invariance
 is violated. As a result the time odd terms also appear. The RMF calculations 
 incorporating  correctly this violation of time invariance are involved  \cite{AFA.10} and 
are beyond the scope 
of the present work which is mainly geared to the calculations  of  $\alpha$ - decay half lives 
of these new isotopes. 
To overcome this difficulty we use the "tagging" prescription. The level (levels) to be 
tagged for the last odd (odd-odd) nucleon are guided by the corresponding 
experimental data or by the results of the deformed RMF calculations for 
neighbouring nuclei. The  "tagging" here means assigning a fixed occupancy to the tagged 
level(s) through out the iterative RMF calculations.
The left over even number of neutrons and even number 
of protons then preserve the time reversal invariance and the calculation 
then proceeds in the conventional manner.  The details can be found in Refs. 
\cite{YKG.90,BHA.11}.

The  $\alpha$- decay half-lives are calculated in the WKB approximation which 
requires the $\alpha$ - daughter nucleus potential (V$_{\alpha D}$) in addition 
to its $Q$ - value. The former is calculated using the double folding procedure, 
by folding the effective nucleon – nucleon potential (M3Y interaction with an 
extra delta function pseudo potential to incorporate the exchange effects)
\cite{SAT.79,AKC.86,DTK.94,DNB.02,GAM.05,GAM.03,GAM.05a}
with RMF density distributions of the daughter nucleus and that of the 
$\alpha$ particle. It is known that the calculated $\alpha$ - decay half-lives 
are very sensitive to its $Q$ - value. Even a few hundred keV difference in 
$Q$ – value can change the calculated half-lives by a few orders of magnitude.
Therefore, accurate (close to the experimental)  Q - values must be used in the  calculation
of decay half lives. 

\section{Results and Discussion}

The calculated total nuclear binding energyies (BE)   along with 
the corresponding  point $\it{rms}$ proton $r_p$  radius of the nuclear density 
distributions, neutron skin ($r_n-r_p$)   and the quadrupole deformation 
parameter $\beta$  are listed in Table I  labeled as  NL3*. Recently, extensive Relativistic Continuum 
Hartree - Bogoliubov (RCHB) calculations  using the uccessful relativistic energy density functional
PC-PK1 (\cite{ZHA.10}), have been reported and the results are presented in \cite{XIA.17}. 
These spherical calculations properly treat the time odd terms  which appear for odd/odd-odd nuclei 
and use blocking procedure for  the relevant tagged state(s) in the calculation. 
The RCHB euations  are solved in the coordinate space (for details see  \cite{ZHA.10,XIA.17}. 
 The relevant results are also
  presented in the same table labelled ad PC-PK1 for comparison and discussion.
Also, the  binding energies BE taken from the recent compilation  
by Wang {\it et al.} \cite{WAN.12} labeled as AU along with the corresponding 
values obtained by using the mass formula of Bhagwat \cite{BHA.14} denoted 
by AB are also shown in the same table. It is to be noted that the AU mostly 
quotes the available experimental values which in fact are almost identical 
(the maximum  deviations are $\sim$  0.5 MeV) to the corresponding AB values. The AB  
masses of nuclei which are yet to be measured can also be obtained.

 The inspection of Table I reveals that the calculated values 
of BE are in good agreement  with the experiment as expected.  
The PC-PK1 values are  considerably,  in better agreement with the experiment. The average deviation 
 for NL3* (PC-PK1)  is $\sim$ 0.6\% (0.1\%).

Both NL3* and PC-PK1  calculated   point $\it{rms}$ proton $r_p$  radius of the nuclear density 
distributions and the  neutron skin ($r_n-r_p$) are very close  to each other, the differences appear only at
 second decimal place of fermi.
 For example the maximum deviation for point $\it{rms}$ proton   radius ($r_p$  is $\sim$ 0.1 fermi while for
 neutron skin $r_n-r_p$ the maximum deviation is  $\sim$ 0.3 fermi.

The calculated NL3*  quadrupole deformation 
parameter $\beta$ values are very similar to the corresponding values of of M\"oller and Nix \cite{MOL.95} for most
of the nuclei, except for the nucleus $^{229}$Am.  Further, most of the 
calculated  NL3* $\beta$ values are small except for $^{233}$Bk where its  value is 0.218, indicating that most 
of these nuclei are close to spherical.

The experimental charge radii  are available \cite{ANG.04} for the following 
four nuclei in this region and  are listed in Table II.  Clearly, both NL3* and PC-KC1 
calculation (RMF) reproduces the corresponding experimental values very well.
 The  deviations appear only in third decimal place of fermi. 

\begin{table}[htb]
\begin{center}
\caption{The calculated (RMF) charge radii along with the corresponding 
experimental values (Expt.) taken from \cite{ANG.04}.}
\addtolength{\tabcolsep}{4pt}
\begin{tabular}{cccc} \hline 
      &  \multicolumn{3}{c}{$r_c$ (fm)}  \\
Nucleus &  NL3*   &  PC-PK1  &  Expt.  \\ \hline \vspace{3pt}
$^{213}_{\,\,87}$Fr$_{126}$ & 5.595  & 5.601 & 5.598 \\\vspace{3pt}
$^{208}_{\,\,87}$Fr$_{121}$ & 5.577  & 5.580 & 5.573 \\\vspace{3pt}
$^{207}_{\,\,87}$Fr$_{120}$ & 5.573  & 5.576 & 5.572 \\\vspace{3pt}
$^{208}_{\,\,88}$Ra$_{120}$ & 5.590  & 5.592 & 5.585 \\\hline \vspace{3pt} 
\end{tabular}
\end{center}
\end{table}

\begin{table}[htb]
\begin{center}
\caption{The $Q$ - values for $\alpha$ - decay chains (for details see text).}
\addtolength{\tabcolsep}{4pt}
\begin{tabular}{ccccc} \hline 
      &  \multicolumn{4}{c}{$Q$ - value (MeV)}  \\
Nucleus &  NL3*  & PC-PK1 &   AB  &  Expt.  \\ \hline \vspace{3pt}
$^{233}_{\,\,97}$Bk$_{136}$ & 8.56 &  11.07  &  8.44 &   7.90    \\\vspace{3pt}
$^{229}_{\,\,95}$Am$_{134}$ & 9.27 &  11.29  &  8.51 &   8.14     \\ \vspace{3pt} 
$^{225}_{\,\,93}$Np$_{132}$ & 9.82 &  11.21  &  8.74 &   8.79$^{*}$      \\\vspace{3pt}
$^{221}_{\,\,91}$Pa$_{130}$ & 9.31 &  10.34  &  9.10 &   9.25$^{*}$ \\\vspace{3pt}
$^{217}_{\,\,89}$Ac$_{128}$ & 9.73 &  11.06  &  9.53 &   9.83$^{*}$ \\\vspace{3pt}
$^{213}_{\,\,87}$Fr$_{126}$ & 4.31 &   4.56  &  6.75 &   6.88       \\\vspace{3pt}
                        &      &         &       &              \\\vspace{3pt}
$^{223}_{\,\,95}$Am$_{128}$ &11.58 &  13.62  & 10.83  & \\\vspace{3pt}
$^{219}_{\,\,93}$Np$_{126}$ & 8.35 &   8.12  &  9.10  &  8.98$^{*}$\\ \vspace{3pt} 
$^{215}_{\,\,91}$Pa$_{124}$ & 6.34 &   5.67  &  8.35  &  8.23      \\\vspace{3pt}
$^{211}_{\,\,89}$Ac$_{122}$ & 5.47 &   5.13  &  7.61  &  7.62$^{*}$  \\\vspace{3pt}
$^{207}_{\,\,87}$Fr$_{120}$ & 5.21 &   5.16  &  6.90  &  6.89$^{*}$  \\\vspace{3pt}
                        &      &         &       &              \\\vspace{3pt}
$^{216}_{\,\,92}$U$_{124}$  & 6.31 &  6.10   & 8.79  & 8.49\\\vspace{3pt}
$^{212}_{\,\,90}$Th$_{122}$ & 6.21 &  5.40   & 8.07  & 7.98\\\vspace{3pt} 
$^{208}_{\,\,87}$Fr$_{121}$ & 4.78 &  4.70   & 6.82  & 6.74\\ \hline  \vspace{3pt} 
\end{tabular}
\end{center}
\end{table}
\begin{table*}[htb]
\begin{center}
\caption{The  estimates of the $\alpha$-- decay half-lives (in
seconds) along with the  corresponding experimental values. For details
see text.}
\begin{tabular}{ccccccc} \hline  
      &  \multicolumn{4}{c}{WKB-with Q-values} &  Pheno & Measured [ref-1] \\\hline  
Nucleus &  NL3*   &   PC-PK1 &   AB  &  Expt.  & Pheno  & Expt[ref-1] \\\hline \vspace{3pt}
$^{233}_{\,\,97}$Bk$_{136}$  &  1.15 $\times$10$^{-1}$  & 1.08                            &  0.28  & 19.0  & 22.58            &  21$^{+48}_{\,-17}$ \\\vspace{3pt}
$^{229}_{\,\,95}$Am$_{134}$ &  1.93$\times$10$^{-4}$ & 1.14                      &  0.03                                          &   0.49                                    &   0.62                                           &
3.7$^{+10.4}_{\,\,-2.2}$             \\\vspace{3pt} 
$^{225}_{\,\,93}$Np$_{132}$ &  1.58$\times$10$^{-6}$   & 68.10                       &  1.20$\times$10$^{-3}$  &  8.63$\times$10$^{-4}$   &  1.23$\times$10$^{-3}$  &                                            
3.6$^{+10.2}_{\,\,-2.2}\times$10$^{-3}$ \\\vspace{3pt}
$^{221}_{\,\,91}$Pa$_{130}$ &  6.53$\times$10$^{-6}$   & 4.15$\times$10$^{-9}$ &  2.43$\times$10$^{-5}$ &  9.68$\times$10$^{-6}$  &   1.45$\times$10$^{-5}$     &                                 
5.9$^{+1.7}_{\,\,-1.7}\times$10$^{-6}$ $^{*}$  \\\vspace{3pt}
$^{217}_{\,\,89}$Ac$_{128}$ &  5.86$\times$10$^{-7}$   & 4.83$\times$10$^{-8}$& 4.20$\times$10$^{-7}$ &  8.09$\times$10$^{-8}$    &   1.23$\times$10$^{-7}$  &                                 
 6.9$^{+0.4}_{\,\,-0.4}\times$10$^{-8}$ $^{*}$  \\\vspace{3pt}
$^{213}_{\,\,87}$Fr$_{126}$ &  2.81$\times$10$^{15}$   & 14.70                                  &  38.20                                   &  11.20                           & 12.20                                             &
 63$^{+177}_{\,-38}$  \\\vspace{3pt}                                           
                                                       &                                                  &                                  &                                                         &                                          &                                              &    \\\vspace{3pt}                                        
$^{223}_{\,\,95}$Am$_{128}$ &  1.58$\times$10$^{-9}$  &  4.08$\times$10$^{-9}$ &  5.32$\times$10$^{-8}$ &  3.39$\times$10$^{-8}$  &   7.31$\times$10$^{-8}$    & 
5.2$^{+12.0}_{\,\,-4.4}\times$10$^{-3}$   \\\vspace{3pt}
$^{219}_{\,\,93}$Np$_{126}$ &  2.82$\times$10$^{-2}$   & 2.32$\times$10$^{-3}$&  1.79$\times$10$^{-4}$ &   1.16$\times$10$^{-4}$   & 1.47$\times$10$^{-4}$     &
           \\\vspace{3pt} 
$^{215}_{\,\,91}$Pa$_{124}$ &  1.98$\times$10$^{5}$    & 0.14                                      &  5.32$\times$10$^{-3}$ &   1.21$\times$10$^{-2}$   &  1.31$\times$10$^{-2}$        &
6.0$^{+13.8}_{\,\,-5.0}\times$10$^{-3}$ \\\vspace{3pt}
$^{211}_{\,\,89}$Ac$_{122}$ &  6.90$\times$10$^{8}$    & 0.22                                    &  0.23                                        &   0.20                               &    0.23                                        &
 7.80$^{+179.0}_{\,\,-64.0}\times$10$^{-2}$ \\\vspace{3pt}
$^{207}_{\,\,87}$Fr$_{120}$ &  2.04$\times$10$^{9}$    & 6.43$\times$10$^{2}$   &  13.80                                     &   14.20                            &  14.46                                        &
 1.7$^{+3.9}_{\,\,-1.4}$   \\\vspace{3pt}
                                                      &                                                  &                                    &                                                             &                                       &                                                &        \\\vspace{3pt} 
$^{216}_{\,\,92}$U$_{124}$  & 9.44$\times$10$^{5}$     & 3.85$\times$10$^{-2}$  &  6.59$\times$10$^{-4}$  &  4.85$\times$10$^{-3}$   &  2.68$\times$10$^{-3}$        &
3.8$^{+8.8}_{\,\,-3.2}\times$10$^{-3}$ \\\vspace{3pt}  
$^{212}_{\,\,90}$Th$_{122}$  & 3.10$\times$10$^{5}$    & 0.04                         &  0.02                                                 &   0.03                               &   2.31$\times$10$^{-2}$         &
0.17$^{+0.40}_{\,\,-0.14}$ \\\vspace{3pt}
$^{208}_{\,\,87}$Fr$_{121}$  & 1.17$\times$10$^{12}$    & 1.06$\times$10$^{2}$   &  26.10                                &    55.00                           &  85.09                                              &
   51$^{+117}_{\,\,-42}$   \\\hline \vspace{3pt}
\end{tabular}
\end{center}
\end{table*}

The calculated (NL3*) $Q$ values of $\alpha$ - decay of the new isotopes and their
 daughter nuclei 
are listed in Table III.  The corresponding PK-PC1 values taken from   \cite{XIA.17} 
 and with that obtained by using the mass formula of Bhagwat \cite{BHA.14} 
labeled as  AB along with the corresponding available experimental values 
(Expt.) \cite{DEV.15} are listed in the same table for  comparison. The experimental
 Q  value of the
 new isotope $^{223}$Am is not available.
 The corresponding values of the rest of the nuclei taken from the live
 chart \cite{iaea-live}
 marked by superscript ($*$) are also shown in the same table for completeness.  
It is clear 
from the table that all the three 
(NL3*, PC-PK1 and AB) $Q$ values are similar and reasonably 
agree with the experiment.  Quantitatively, the deviations rise  even up to 3 MeV in some cases.
This type of agreement is considered to be quite 
acceptable in view of the fact that the $Q$ – value is the difference 
between the binding energies (BE) of the parent nucleus and the daughter 
nucleus plus that of $\alpha$ particle. The BE themselves are very large. A small 
error in one may easily upset the quality of the obtained agreement. 
 At the finer level  it is observed that – AB values  are closest to the 
Expt. .   The NL3*  Q  values though are similar to those 
of AB  and the Expt. values but differ substantially ($>$1 MeV) in several cases 
(e.g. for    $^{229}_{\,\, 95}$Am,  $^{225}_{\,\, 93}$Np,  $^{213}_{\,\, 87}$Fr, $^{215}_{\,\, 91}$Pa, 
$^{211}_{\,\, 89}$Ac, 
$^{207}_{\,\,87}$Fr, $^{216}_{\,\, 92}$U, $^{212}_{\,\, 90}$Th , and $^{208}_{\,\, 87}$Fr).
Similar deviations, for the  PC-PK1 case  are relatively larger as compaed to that of NL3*. As we shall 
see, these differences play a very crucial role in the calculation of 
the corresponding $\alpha$ -  decay half-lives. We end this discussion by restating that the Relativistic 
Mean Field (Relativistic eenergy density functional) theories provide accurate and reliable description of
nuclear ground state properties for nuclei spread over the entire periodic table.

\begin{figure}
\includegraphics[width=0.45\textwidth]{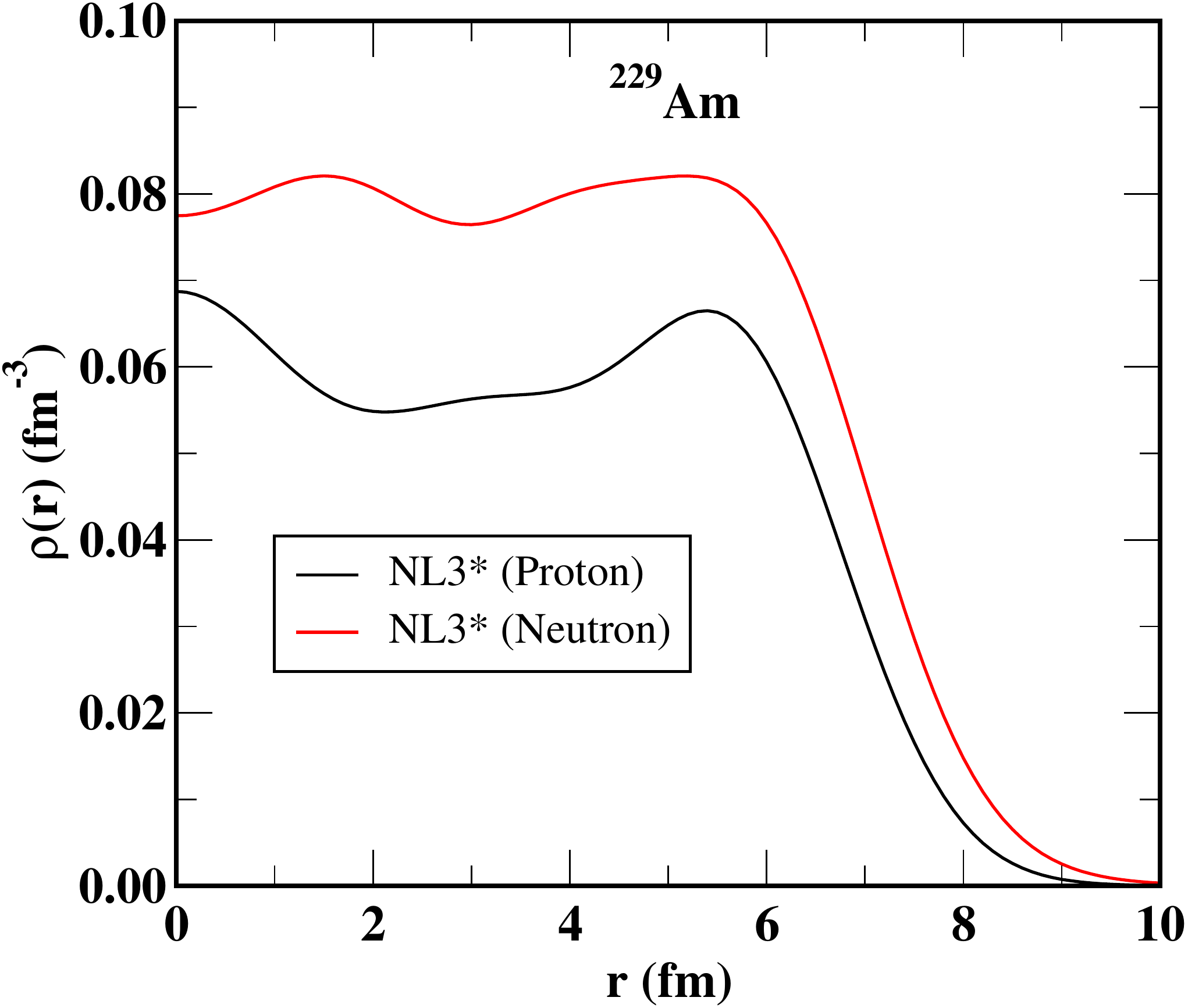}
\caption{ Calculated RMF Nucleon density distribution using N3* set of parameters .}
\label{fig2}
\end{figure}

As stated before, the  $\alpha$ - decay half-lives are calculated in 
the WKB approximation. The required  $\alpha$ - daughter nucleus potential  
(V$_{\alpha D}$)  is calculated  by folding the effective nucleon – nucleon 
potential (M3Y interaction + pseudo potential to incorporate the exchange 
effects) with RMF density distributions of the daughter nucleus and that 
of the $\alpha$ - particle. 

It is known (e.g. see \cite{YKG.90})  that the calculated 
RMF nucleon density distributions give good account of the experiment.  The calculated
RMF nucleon (both proton and neutron) density distributions obtained by using NL3* 
Lagrangian parameters are shown in Fig. 2 for $^{229}$Am.   Very small oscillations 
appearing in some of the calculated density distributions at short distances  (see Fig. 2) 
smear out during the folding process, resulting into  a smooth  $\alpha$ - daughter nucleus
 potential  (V$_{\alpha D}$). As an illustration   the calculated   $\alpha$ - $^{229}$Am -  potential  
 is shown in figure 3.  It is expected that such  calculated V$_{\alpha D}$ potentials 
are reliable and can be used with confidence in the WKB framework for the calculation 
of $\alpha$ - decay half lives.

\begin{figure}
\includegraphics[width=0.45\textwidth]{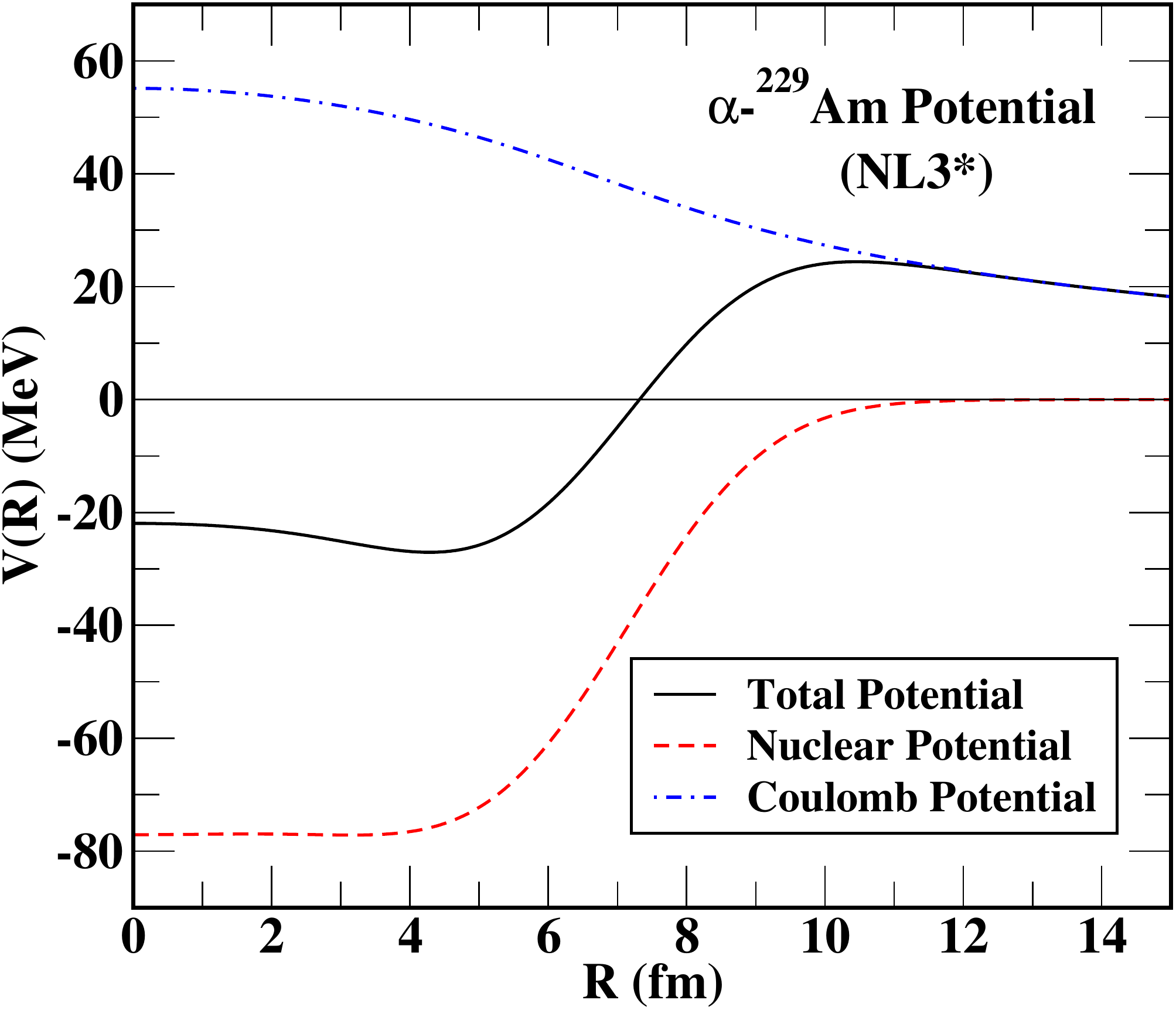}
\caption{ Calculated  $^{229}$Am - $\alpha$ potential (V$_{\alpha D}$) .}
\label{fig3}
\end{figure}

Several analytical (phenomenological expressions  involving number of  adjustable parameters)
 for the calculation of $\alpha$ - decay half lives, mostly based on the original viola-Seaborg formula 
(\cite{VIO.66}) are also available in the literature. For  completeness, we have used one 
of the recent expression    reported by Dasgupta  - Schubert and Reyes (\cite{DAS.07}) is given below
for $\alpha$ - decay half lives $T_{1/2}$ in Seconds: 

\begin{equation}
\log_{10} \left(T_{1/2}\right)  = a + b A^{1/6} Z^{1/2} + c Z Q^{-1/2}
\end{equation}
with parameters:\\
$ a =-25.31, b =-1.1629, c = 1.5864 $ for even- Z, even- N \\
$ a =-25.65, b =-1.0859, c = 1.5848$ for even- Z, odd- N \\
$ a =-25.68, b =-1.1423, c = 1.5920$ for odd- Z, odd- N \\
$ a =-29.48, b =-1.1130, c = 1.6970$ for odd- Z, odd- N \\

We have calculated the relevant $\alpha$ - decay half lives using this 
expression (Eq.1), for comparison and discussion.

It is known that the calculated   WKB $\alpha$ -  decay half-lives are very.
sensitive to its respective $Q$ - values. Even a few hundred keV difference 
in $Q$ - value can change calculated half-lives by a few orders of magnitude 
(see Ref. \cite{GAM.05}). The results are displayed in Table IV. The symbols 
NL3*, PC-PK1,  AB and Expt. correspond to the calculated WKB  $\alpha$- decay half-lives using  their respective $Q$ - values (listed in Table III). 

 The measured values
reported in \cite{DEV.15} and listed in Table IV under the header Expt.[ref-1]. have very
 large errors (mostly statistical). It is to be noted that two values of the measured
 $\alpha$- decay  half-lives
are quoted in Table I  of Ref. \cite{DEV.15} for
$^{229}$Am,$^{225}$Np and $^{213}$Fr nuclei. Therefore, an average of these two measured
values are listed  as experimental (Expt.[ref-1]) values  in Table IV for each of
these nuclei. The values of the half-lives for the known isotopes which have not been measured in
\cite{DEV.15} taken from the live chart \cite{iaea-live} are also shown with the superscript $^*$. 
The $\alpha$ - decay half-life for the nucleus $^{219}$Np has not been reported in \cite{DEV.15} 
and also is not available in the live chart. The corresponding  $\alpha$- decay half-lives, calculated 
 using the phenomenological expression (Eq.1) with the experimental (Expt. listed in Table III) Q-values
are also shown in Table IV under the header Pheno, for comparison and discussion.

 It is clear from the Table IV
that the calculated half-lives (RMF, MN, AB and Expt.)  using their respective  $Q$ - values listed in Table III,
differ widely among themselves indicating high sensitivity on the $Q$ - values 
used, as expected. It is intetresting to note that  all the estimated decay half lives 
( NL3*, PC-PK1,  AB and 
also Expt.  and Pheno (where AB-Q values are used))   obtained by using the corresponding $Q$ values listed
 in Table III for   $^{223}$Am  
 are  very small  ($\sim $   $\times$10$^{-8}$ seconds  while the corresponding  experimental value (Expt.[ref-1]) 
is comparitively very large ($\sim $ $\times$10$^{-3}$ seconds. This case was marked "pile up"  by the authors of  \cite{DEV.15}, This  then indicates the necessity for repeating their procedure for extracting this value of 
half life for  $^{223}$Am.  

The calculated (NL3* and PC-PK1)  half lives differ considerably,  from the corresponding experimental values in most of the cases,
 either grossely underesatimating  or overestimating the experimental values.  This again is a reflection of the high 
sensitivity on the $Q$ - values used in the calculation of half lives.  Therefore the calculated values of the  half lives  for both  NL3*
 and PC-PK1 are not reliable, mainly due to the fact that their respective $Q$ values differ from the experiment and therefore
are not accurate enough to be used in the WKB calculation of half lives.

The results under the header AB where the used $Q$ - values 
which well represent the experimental $Q$ - values, agree well (within a factor of $\sim$8) with
the corresponding experimental values except for $^{233}$Bk  and $^{229}$Am  where it considerably
 underestimaes ($\sim $ by a factor of 100) the experimental value.  
 On the other hand  the  corresponding  Expt. and Pheno values obtained by using  the experimental 
$Q$ values  almost reproduces the experiment,  indicating that the  
said deficiency  can be   rectified 
by using  accurate (experimental or very close to the experimental) $Q$ values.
For the rest of the nuclei, the values of  half lives presented in Table IV for all the three AB, Pheno and Expt., are very similar to each other and reasonably 
agree with the corresponding experimental (expt.[ref-1])
  values.   The  calculated AB and Pheno half lives are  very close  (with in a factor of 4) to each other   except for $^{233}$Bk and $^{229}$Am and also  reasonably agree with the experiment. 
The Expt. values improve this agreement with the experiment further as is evident from the values listed in Table IV.

 From the above discussion 
it may be concluded that the microscopic estimates AB,  and Expt.  and so also Phrno of the $\alpha$ - decay half-lives using the 
corresponding experimental (where available) ground state Q values represent a fair estimate of
the corresponding experimental (measured and yet to be measured) values. 

The experimental (Expt.[ref-1]) values of half lives listed in Table IV range from seconds to 
 $\sim $   $\times$10$^{-6}$ seconds, besides having large errors,  
in view of this observation 
the agreement  with in a factor of 10 with the experiment  is acceptable  in such studies
in this nuclear region. 
Therefore the agreement obtained here  between the calculed and the experimental half lives is 
 remarkable indeed.

 It is  strongly advocated  that the use of the experimental ground state
 $Q$ - values is essential in the microscopic calculation of the half-lives to obtain 
good description of the experiment.  

Most of the nuclei considered here are unstable and decay through various decay
modes  like  $\alpha$ - emission, fission, electron capture etc.. Usaually,
one or two decay modes dominate. A complete microscopic description of complex 
fission process is still lacking, though some semi microscopic model descriptions 
do exist (e.g. see \cite{SGN.69,BRA.72,GR.75,SOB.69,SMO.95,MUT.03,ZBE.14}). 
The detailed calculations are quite involved and are beyond the scope of the present work.

\section{Summary and Conclusion}
The ground state properties of the recently reported new isotopes  
$^{216}_{\,\,92}$U, $^{223}_{\,\,95}$Am,
$^{219}_{\,\,93}$Np, $^{229}_{\,\,95}$Am and $^{233}_{\,\,97}$Bk
 along with the elements appearing in their respective $\alpha$ - decay chains are calculated 
in the frame work of the relativistic mean field (RMF) theory. The calculated binding 
energy per nucleon (BE/A), the neutron (proton) point radii $r_n$ ($r_p$) of the 
RMF density distributions, the neutron skin ($r_n - r_p$), the  mass and the charge radii, 
the quadrupole deformation parameter $\beta$ and the corresponding
 $Q$ - values  of $\alpha$ - decay, compare well with the corresponding experimental values 
(where available).  The $\alpha$ - decay half-lives are calculated using the
WKB approximation which requires the corresponding $Q$ - values and the parent - 
daughter potential.  The latter is obtained in the double folding procedure 
($t\rho\rho$ - approximation). Though the calculated RMF and MN $Q$ - values are 
close to the corresponding experimental values, the small differences between them and the 
experimental values , play 
a crucial role and may even change the  calculated half-lives by a few orders 
of magnitude.  Therefore,  the present RMF and MN $Q$ values are not accurate enough to be used in the 
WKB calculation of  the  $\alpha$ - decay half lives.

It is strongly emhasized that the use of the accurate( expterimental or very close to the experimental)  $Q$ values  
 is crucial  in  the microscopic as well as phenomenological,  calculations of the $\alpha$ - decay half-lives. 
\begin{acknowledgments}
AB acknowledges partial financial support from the Department of Science and 
Technology, Govt. of India and the Swedish National Research Council (VR) 
(through grant No. DST/INT/SWD/VR/P-04/2014). 
H.M. Devaraja, acknowledges financial support from the LOEWE program.
\end{acknowledgments}


\begin{thebibliography}{99}
\bibitem{DEV.15} H. M. Devaraja {\it et al.}, Phys. Lett. B {\bf 748}, 
199 (2015).

\bibitem{MUN.79} G. M\"unzenberg {\it et al.}, Nucl. Instrum. Methods  {\bf 161}, 65 (1979).


\bibitem{PGR.89} P. -G. Reinhardt, Rep. Prog. Phys. {\bf 52}, 439 (1989)
and references cited therein.         
\bibitem{YKG.90} Y. K. Gambhir, P. Ring and A. Thimet, Ann. Phys. (NY) 
{\bf 198}, 132 (1990).
\bibitem{PRI.96} P. Ring, Prog. Part. Nucl. Phys. {\bf 37}, 193 (1996) and 
references cited therein.                       
\bibitem{VRE.05} D. Vretenar, A. V. Afanasjev, G. A. Lalazissis and 
P. Ring, Phys. Rep. {\bf 409}, 101 (2005) and references cited therein.
\bibitem{BHA.11} A. Bhagwat and Y. K. Gambhir,  Int. J. Mod. Phys. 
E {\bf 20}, 1663 (2011).
\bibitem{AGB.14} S. E. Agbemava, A. V. Afanasjev, D. Ray and P. Ring, 
Phys. Rev. C {\bf 89}, 054320 (2014).
\bibitem{ZHA.05} W. Zhang {\it et al.}, Nucl. Phys. A {\bf 753}, 
106 (2005).
\bibitem{LAL.09} G. A. Lalazissis,  S. Karatzikos, R. fossion, D. Pena Arteaga,
 A. V. Afanasjev   and P. Ring, Phys. Lett. B{\bf 671}, 36 (2009).
\bibitem{LAL.97} G. A. Lalazissis, J. K\"onig and P. Ring, Phys. Rev. C
{\bf 55}, 540 (1997).
\bibitem{BER.84} J. F. Berger, M. Girod and D. Gogny, Nucl. Phys. A {\bf 428},
23 (1984).
\bibitem{GON.96} T. Gonzalez-Llarena, J. L. Egido, G. A. Lalazissis and P. Ring,
Phys. Lett. B {\bf 379}, 13  (1996). 
\bibitem{AFA.10}  A. V. Afanasjev and H. Abusara, 
Phys. Rev. C {\bf 81}, 014309 (2010).
\bibitem{SAT.79} G. R. Satchler and W. G. Love, Phys. Reports {\bf 55} 183, (1979). 
\bibitem{AKC.86} A. K. Choudhuri, Nucl. Phys. A {\bf 449}, 243 (1986).
\bibitem{DTK.94} D. T. Khoa, W. von Oertzen and H. G. Bohlen, 
Phys. Rev. C {\bf 49} 1652, (1994).  
\bibitem{DNB.02} D. N. Basu, Phys. Rev. C {\bf 66}, 027601 (2002) and 
references cited therein.                 
\bibitem{GAM.05} Y. K. Gambhir, A. Bhagwat and M. Gupta, 
Ann. Phys. (N.Y.) {\bf  320}, 429 (2005).
\bibitem{GAM.03} Y. K. Gambhir, A. Bhagwat , M. Gupta and Arun K. Jain,
Phys. Rev. C {\bf 68},  044316 (2003).
\bibitem{GAM.05a} Y. K. Gambhir, A. Bhagwat and M. Gupta, 
Phys. Rev. C {\bf 71}, 037301 (2005). 


\bibitem{ZHA.10}   P.W. Zhao, Z.P. Li, M. Yao and J. Meng,
Phys. Rev. C {\bf 82}, 054319 (2010). 
\bibitem{XIA.17} X.W. Xia, Y. Lim, P.W. Zhao,  H.J. Liang, X.Y. Qu, Y. Chen, H. Lin, L.F. Zhang, 
S.Q. Zhang, Y. Kim and J. Meng, arXiv:1704.08906v2[nucl-th], 14 Sept. 2017.
Chin. Phys. Lett.  {\bf 29}, 042101  (2012). 


\bibitem{WAN.12} M. Wang {\it et al.}, Chinese Phys. C {\bf 36}, 1287 (2012), and 
{\bf 36}, 1603 (2012)  and {\bf 41}, 03003 (2017).
\bibitem{BHA.14} A. Bhagwat, Phys. Rev. C {\bf 90}, 064306 (2014). 

\bibitem{ANG.04} I. Angeli and K. P. Marinova, {\it At. Data Nucl. Data Tables} 
{\bf 99} (2013) 69. 
\bibitem{MOL.97} P. M\"oller, J. R. Nix and K. -L. Kratz, 
At. Data Nucl. Data Tables {\bf 66}, 131 (1997). 
\bibitem{iaea-live} https://www-nds.iaea.org/relnsd/vcharthtml/ VChartHTML.html.

\bibitem{VIO.66} V. E. Viola and G. T. Seaborg, Journal of inorganic and
 nuclear chemistry {\bf 28}, 697 1966.
\bibitem{DAS.07} N. Dasgupta - Schubert, M.A. Reyes,   
At. Data Nucl. Data Tables {\bf 93}, 90 (2007). 



\bibitem{DAS.07} N. Dasgupta - Schubert, M.A. Reyes,   
At. Data Nucl. Data Tables {\bf 93}, 90 (2007). 

\bibitem{SGN.69} S. G. Nilsson {\it et al.}, Nucl. Phys. A {\bf 131}, 1 (1969).
\bibitem{BRA.72} M. Brack {\it et al.}, Rev. Mod. Phys. {\bf 44}, 320 (1972).
\bibitem{GR.75} G. Ripka, Nuclear Self consistent Fields (1975) (North Holland, New York).
\bibitem{SOB.69} A. Sobiczewski {\it et al.}, Nucl. Phys. A {\bf 131}, 67 (1969).
\bibitem{SMO.95} R. Smolanczuk, J. Skalski, A. Sobiczewski, Phys. Rev. C 
{\bf 52}, 1871 (1995).
\bibitem{MUT.03} I. Mutain, Z. Patyk and  A. Sobiczewski, Acta. Phys. Pol. 
{\bf B34}, 2141 (2003).
\bibitem{ZBE.14} A. Zdeb, M. wardaand K. Pomorski Phys. Scr. {\bf 89}, 054015 (2014).

\end{thebibliography}
\end{document}